\documentclass[prd,longbibliography,amsmath,amssymb]{revtex4}

\usepackage{graphicx}
\usepackage{epsfig}
\usepackage{amssymb,amsmath,amsfonts}
\usepackage[hidelinks]{hyperref}
\usepackage{overpic} 
\usepackage{wasysym}
\usepackage{latexsym}
\usepackage{array,multirow}
\usepackage{verbatim}
\usepackage[normalem]{ulem}
\usepackage{color}
\definecolor{orange}{rgb}{1,0.45,0}
\usepackage{amssymb,amsmath,placeins,psfrag,graphicx,mathtools}
\usepackage{eucal}
\usepackage{tikz}
\usetikzlibrary{arrows}

\newcommand{\beq}{\begin{equation}}
\newcommand{\eeq}{\end{equation}}

\usepackage{slashed}

\begin{document}
\title{Edge states and anomalous SUSY in  (2+1)-dimensional Maxwell Chern-Simons theory  }

\author{Nirmalendu Acharyya}
\email{nirmalendu@iitbbs.ac.in}
\author{Akash Sinha}

\affiliation{School of Basic Sciences, Indian Institute of Technology Bhubaneswar, Jatni, Khurda, Odisha 752050, India}

\date{\today }

\begin{abstract}
In a $(2+1)$-dimensional Maxwell-Chern-Simons theory coupled with a fermion and a scalar, which has $\mathcal{N}=2$ SUSY in absence of the boundary, the insertion of a spatial boundary breaks the supersymmetry. We show that only a subset of the boundary conditions allowed by the self-adjointness of the Hamiltonian can preserve partial $\mathcal{N}=1$ supersymmetry, while for the remaining boundary conditions SUSY is completely broken.  In the latter case, we demonstrate two distinct SUSY-breaking  mechanisms. For some of the SUSY-breaking boundary conditions, the SUSY variation of the action does not vanish which explicitly breaks SUSY.  While for certain other boundary conditions,  despite the invariance of action under SUSY transformations,  unpaired fermionic edge states in the domain of the Hamiltonian leads to an anomalous breaking of the supersymmetry.

\end{abstract}

\maketitle
\section{Introduction} 
Chern-Simons (CS) theory in (2+1)-dimensional manifold has been a subject of interest ever since its conception. On one hand, the topologically massive gauge theory, particularly on a manifold with spatial boundary, itself  has many striking features \cite{Witten:1988hf, Elitzur:1989nr}. On the other hand,  it has played a pivotal role in shaping our understanding of several phenomena in condensed matter systems. In the gapped phases of such systems, the CS theory describes the low-energy effective dynamics of the bulk and the gapless edge-localized excitations \cite{Wen:1995}. The CS theory provides a neat explanation of the integer and fractional quantum Hall effect and  the presence of chiral currents on the edge  \cite{Grivin:1987, Zhang:1988}. Subsequently,  it  has been related to high $T_c$ superconductivity \cite{Laughlin:1988}.  Further,  Abelian CS gauge theory reveals topological order in (2+1)-dimensional quantum spin liquids and superconductors \cite{Kalmeyer:1987, Hansson:2004, Kou:2008}.  Apart from that, it emerges in the low energy description of interacting Dirac fermions in (2+1) dimensions \cite{Palumbo:2013rb}.   A recent  study elucidates  the existence of topological electromagnetic phases in the Abelian Maxwell-Chern-Simons theory \cite{Mechelen:2020}.

In this article, we consider a topologically massive Maxwell-CS theory along with a real scalar field and Dirac fermion in a (2+1)-dimensional manifold $M$ with a spatial boundary $\partial M$. In absence of the boundary, this system exhibits $\mathcal{N}=2$ supersymmetry (SUSY) in the full (2+1)-dimensional spacetime. Here, we investigate the fate of SUSY on insertion of a spatial boundary.  Though the experimental observations for SUSY as a fundamental theory is yet to come, a recent study found that spacetime supersymmetry emerges naturally in an effective low-energy description of quantum phase transitions at the boundary of topological superconductors and insulators \cite{Grover:2013rc}. This led to a revival of interest in supersymmetric effective theories of condensed matter systems \cite{Rahmani:2015,Hsieh:2016emq,Prakash:2020krs, Jian:2016zll,Ma:2021dua} and in particular, (2+1)-dimensional CS theory \cite{Bae:2021lvk}. 

Insertion of a boundary $\partial M$, in general, reduces the symmetries of the system. Therefore, it is natural to ask whether SUSY in the manifold $M$ can be obtained by truncation of  $\mathcal{N}=2$ super-Maxwell-CS theory in the entire spacetime. As expected \cite{Belyaev:2008xk, Okazaki:2013kaa, Acharyya:2015swa}, here we will demonstrate  that supersymmetry in the super-Maxwell-CS theory can be partially preserved only when certain specific boundary conditions are imposed on the fields. Thus, the boundary conditions assume a crucial role in determining the fate of the supersymmetry and it is necessary to classify them as SUSY-preserving or SUSY-breaking. Of course, the boundary conditions on the fields cannot be chosen arbitrarily.  A boundary condition should be such that the fields belong to self-adjoint domains of the Hamiltonian. This ensures the real energy eigenvalues and the field can be expanded in the basis of the eigenfunction of the Hamiltonian and quantized.

Here, we obtain the set of all allowed boundary conditions on the gauge fields, scalar field and the fermion and demonstrate that only a small subset of it can preserve SUSY, at least partially. Imposing  boundary conditions outside this subset breaks SUSY completely.  Further, we show that there are two distinct mechanisms of the SUSY breaking in these cases. For some of these SUSY-breaking boundary conditions, the supersymmetric variation of the classical action does not vanish, rendering them incompatible with SUSY. 

On the other hand, we show that there is another class of SUSY-breaking boundary conditions which breaks supersymmetry despite the invariance of the action under any SUSY transformation. In these scenarios,  we demonstrate the existence of edge-localized counter-propagating fermionic excitations with non-zero energy which has no bosonic counterpart. As a result, the action of the supercharges on these fermionic states changes the domain (similar to \cite{Esteve:1986db}) and  supersymmetry is anomalously broken in the quantum theory.


\section{Chern-Simons gauge theory  in a $(2+1)$-dimensional manifold} 
In the full (2+1)-dimensional spacetime,  $U(1)$ gauge theory with a Dirac fermion $\psi$ and real scalar field $\phi$ described by the Lagrangian density $ \left( \mathcal{L}_{M} + \mathcal{L}_{CS}\right) $ with
\begin{eqnarray}
\begin{array}{l}
\mathcal{L}_{M} = - \frac{1}{4} F^{\mu\nu} F_{\mu \nu} + \frac{1}{2}\partial_\mu \phi \partial^\mu \phi + \frac{1}{2}D^2+i \bar{\psi} \gamma^\mu \partial_\mu \psi, \\ \\
\mathcal{L}_{CS} = m \left(- \frac{1}{2} \epsilon^{\mu \nu \rho} A_\mu \partial_\nu A_\rho + \bar{\psi}{\psi} - \phi D\right)
\end{array} \label{Lagrangian_full_space} 
\end{eqnarray}
has $\mathcal{N}=2$ supersymmetry \cite{Hook:2013yda}. Here, $A_\mu$'s are the gauge fields, $F_{\mu \nu} =\partial_\mu A_\nu -\partial_\nu A_\mu$ and  $m$ denotes the Chern-Simons coupling constant. $\psi$ is a 2-component spinor: $\psi = (\psi_1\,\,\,\, \psi_2)^T$ and  $D$ is a real auxiliary scalar. 

The Dirac-$\gamma$ matrices  satisfy
\begin{eqnarray}
&& \{\gamma^\mu, \gamma^\nu\} = 2 \eta^{\mu\nu}, \quad\gamma^{\mu\nu} \equiv [\gamma^\mu, \gamma^\nu] = -2 i \epsilon^{\mu\nu\rho} \gamma_\rho, \nonumber \\
&& \gamma^\mu \gamma^\nu \gamma^\rho = \eta^{\nu \rho} \gamma^\mu -  \eta^{\mu \rho} \gamma^\nu +  \eta^{\mu \nu} \gamma^\rho - i \epsilon^{\mu \nu \rho} 
\end{eqnarray}
where $\eta=\text{diag}(1,-1,-1)$. We can choose $\gamma^\mu$ in the following representation:  $\gamma^0 = \sigma_2,  \gamma^1= i \sigma_3$ and $ \gamma^2= i \sigma_1$,
where $\sigma_i$'s are the Pauli matrices.

The SUSY transformation are 
 \begin{eqnarray}
 \begin{array}{l}
\delta \phi =i \left( \bar{\epsilon} \psi - \bar{\psi}  \epsilon \right), \,\, \delta \psi = \left( \frac{1}{4} \gamma^{\mu\nu} F_{\mu \nu} - \gamma^\mu \partial_\mu \phi -i D\right)\epsilon , \\ \\
\delta A_\mu = i \left( \bar{\epsilon} \gamma_\mu \psi - \bar{\psi} \gamma_\mu \epsilon \right), \,\, \delta D = \left( \bar{\epsilon} \gamma^\mu \partial_\mu \psi + \partial_\mu \bar{\psi} \gamma^\mu \epsilon\right) 
 \end{array}\label{SUSY_trans_MCS}
\end{eqnarray}
with the supersymmetry parameter $\epsilon = (\epsilon_1\,\,\,\, \epsilon_2)^T$ and $\bar{\epsilon} = \epsilon^\dagger \gamma^0$ 
where $\epsilon_i$'s  Grassmann constants. These SUSY transformations (\ref{SUSY_trans_MCS}) are generated by four supercharges $Q_\alpha$ and $\bar{Q}_\alpha$ with $\alpha =1,2$. 

Here, we consider the same in a  $(2+1)$-dimensional manifold $M =\{x_0, x_1,x_2: x_2\geq 0\}$ with spatial boundary $\partial M$ at $x_2=0$.  Choosing $A_0=0$ gauge, the action is given by 
\begin{eqnarray}
S = \int_M d^3 x \left(\mathcal{L}_{M} + \mathcal{L}_{CS} \right) + S_B
\end{eqnarray}
where $S_B$ are the boundary terms \cite{Asorey:2006pr, Asorey:2008xt} 
\begin{eqnarray}
S_B =-\frac{1}{2} \int_{ \partial M} d^2 x \left(   \phi \partial_2 \phi  -i \bar{\psi} \gamma^2 \psi 
 \right). 
\end{eqnarray}
The boundary terms (analogous to the Gibbons-Hawking term) are required to ensure identical local equations of motion irrespective of boundary conditions on the
fields.

\subsection{Hamiltonian and boundary conditions}

With the gauge field $A_i$ and its conjugate momenta $\Pi_i$,  we define the one-forms (for details, see appendix)
\begin{eqnarray}
A \equiv A_i dx^i, \quad\quad \Pi \equiv \Pi_i dx^i. 
\end{eqnarray}
The electric and magnetic fields are given by  $E = (\Pi - \frac{1}{2} m \ast A)$ and $B = \ast dA$, respectively.  At the classical level, the fields $A_i$ and the momentum $\Pi_i$ satisfy canonical equal-time Poisson brackets, which  in the quantum theory becomes $[A_i(\vec{x},t) , \Pi_j(\vec{y},t)]= i \delta_{ij} \delta^2(\vec{x}-\vec{y})$.

The Hamiltonian is given by $H = H_G + H_s +H_f$ with
\begin{eqnarray}
&& H_G= \frac{1}{2} \int_M d^2x \Big(|E_i|^2 + A_i ( \widehat{ \mathcal{H}}_G A)_i  \Big),\quad\quad \nonumber \\
&& H_s= \frac{1}{2} \int_M d^2x \left(\Pi_\phi^2 +  \phi (\widehat{ \mathcal{H}}_s\phi) \right), \quad\quad \\
&&H_f=  \int_M d^2x \,\, \psi^\dagger (\widehat{ \mathcal{H}}_f \psi ) \nonumber
\end{eqnarray}
where $\Pi_\phi$ is the conjugate momentum of the scalar field $\phi$ and 
\begin{eqnarray}
&
\widehat{ \mathcal{H}}_G \equiv \ast d\ast d, \quad\quad \widehat{ \mathcal{H}}_s\equiv  -\nabla^2 + m^2, \quad\quad 
\nonumber \\
& 
\widehat{ \mathcal{H}}_f \equiv -\left( i \gamma^0 \gamma^i \partial_i + m \gamma^0\right).
\end{eqnarray}
Describing the dynamics of the gauge fields also require the Gauss law \cite{Balachandran:1991dw, Balachandran:1993tm}
\begin{eqnarray}
\mathcal{G}(f) \equiv \int_M d^2 x \,\, (\partial_i f) (\Pi_i + \frac{1}{2} m \epsilon_{ij} A_j) =0
\end{eqnarray}
where $f(x)$ is a test function on $M$ that vanishes on the boundary $\partial M$ and the operator $\mathcal{G}(f)$ vanishes on quantum state vectors in the physical Hilbert space. 


The gauge fields, the scalar and the fermion fields can be expressed in the basis of the eigenfunction of the operators  $\widehat{ \mathcal{H}}_G$,  $\widehat{ \mathcal{H}}_s$ and $\widehat{ \mathcal{H}}_f$, respectively.

To ensure the self-adjointness of  $H_s$, it is necessary that scalar Laplacian $\widehat{ \mathcal{H}}_s$ be self-adjoint \cite{Asorey:2004kk}. With local boundary conditions, this requires that the domain $\mathcal{D}_{\widehat{ \mathcal{H}}_s} = \mathcal{D}_{\widehat{ \mathcal{H}}^\ast_s}$ of $\widehat{ \mathcal{H}}_s$ contains all $\phi \in L^2(M)$ sastisfying
\begin{eqnarray}
\Big[\phi(x) + i \partial_2 \phi(x)\Big]_{\partial M}  = U_s(x) \Big[\phi(x) - i \partial_2 \phi(x)\Big]_{\partial M} 
\end{eqnarray}
with $U_s^\dagger U_s=1$ for all $x \in \partial M$. This leads to either of the following boundary condition on the scalar field: 
\begin{eqnarray}
\begin{array}{lll}
1. \,\,  \text{Neumann boundary condition}: \\ \quad \quad \text{For } U_s= 1, \quad\quad  \partial_2\phi \Big|_{x_2 =0} =0 \\ \\ 
2.\,\,  \text{Dirichlet boundary condition}: \\ \quad \quad  \text{For }U_s = -1, \quad\quad  \phi \Big|_{x_2 =0} =0    \\ \\ 
3.\,\,  \text{Robin boundary condition}:   \\  \quad  \quad\text{For } U_s \neq \pm 1,  \quad\quad    \partial_2\phi \Big|_{x_2 =0} =  \lambda_s \,\, \phi \Big|_{x_2 =0} \\
\end{array} \label{allowed_scalar_bc}
\end{eqnarray}
where $\lambda_s(x) \equiv i(1+U_s(x))^{-1}(1-U_s(x))=\lambda_s^\dagger(x)$  for all $x \in \partial M$.

Similarly, the self-adjointness of $H_G$ requires ensuring 
that the domain $\mathcal{D}_{\widehat{ \mathcal{H}}_G}=\mathcal{D}_{\widehat{ \mathcal{H}}_G^\ast} $ of $\widehat{ \mathcal{H}}_G$ contains all one-forms $A$ with $A_i \in L^2(M)$ satisfying the local boundary conditions  \cite{Balachandran:1993tm, Balachandran:1994vi, Acharyya:2016xaq}:
\begin{eqnarray}
&
 \Big[A_1(x) + i F_{12}(x) \Big]_{\partial M}  = U_G(x) \Big[A_1(x)- i F_{12}(x)\Big]_{\partial M} , 
 \nonumber \\
 &
  U_G^\dagger U_G=1, \quad\quad x \in\partial M. 
\end{eqnarray}
This means, either of the following boundary conditions can be imposed on the gauge fields: 
\begin{eqnarray}
\begin{array}{lll}
1.\,\,  \text{For } U_G = -1 , \quad\quad & A_1 \Big|_{x_2 =0} =0   \\ \\ 
2. \,\, \text{For } U_G = 1, \quad\quad & F_{12}  \Big|_{x_2 =0} =0 \\\\
3.\,\, \text{For } U_G \neq \pm 1,  \quad\quad &  F_{12} \Big|_{x_2 =0} = \lambda_G A_1 \Big|_{x_2 =0}  \\ \\
\end{array} \label{allowed_gauge_bc}
\end{eqnarray}
where $\lambda_G(x) \equiv i(1+U_G(x))^{-1}(1-U_G(x))=\lambda^\dagger_G(x)$ and $0< \lambda_G(x) < \infty $ for all $x \in \partial M$ ensures that $H_G$ is positive semi-definite. 

The magnetic field $F_{12}$ can vanish on the boundary under two circumstances which yields the two distinct boundary conditions: 
\begin{eqnarray}
\begin{array}{lll}
2.a) \quad  \,\, \partial_1 A_2 \Big|_{x_2 =0} =  \partial_2 A_1 \Big|_{x_2 =0}  \\ \\ 
2. b)   \quad \,\,  A_2 \Big|_{x_2 =0} =0 ,\quad   \partial_2 A_1 \Big|_{x_2 =0} =0 .
\end{array} \label{bc_gauge_2} 
\end{eqnarray}

On the other hand, reality of the fermionic Hamiltonian $H_f$ requires  finding the self-adjoint extensions of $\widehat{ \mathcal{H}}_f$. With the projectors $\mathcal{P}_ \pm = \frac{1}{2}(1\pm \gamma^0 \gamma^2)$ on the boundary $\partial M$, we can define 
\begin{eqnarray}
\psi_\pm = \mathcal{P}_ \pm \psi\Big|_{x_2=0}. 
\end{eqnarray}
The self-adjointness of $\widehat{ \mathcal{H}}_f$ requires that  the domain $\mathcal{D}_{\widehat{ \mathcal{H}}_f}$ of $\widehat{ \mathcal{H}}_f$ conatins all $\psi \in W^{1,2}(M) \otimes \mathbb{C}$ satisfying \cite{Asorey:2004kk, Asorey:2013wvh,Asorey:2015sra}
\begin{equation}
 \psi_ + = U_F \gamma^0 \psi_-=0, \,\,\,\,\,U_F^\dagger U_F=1, \,\,\,\,\, [U_F, \gamma^0 \gamma^2]=0. \label{UF_cond_1}
\end{equation}
%
%
The most general $U_F$ satisfying (\ref{UF_cond_1}) is
\begin{equation}
\hspace{-0.1cm} U_F = i\left (\begin{array}{cc}
u_1 & 0 \\
0 & u_2
\end{array}
\right),\,\,\, u_1, u_2 \in \mathbb{C},\,\,\, |u_1|^2=1=|u_2|^2,
\end{equation}
which yields the boundary condition on the fermions
\begin{equation}
 \psi_1\Big|_{x_2=0}  = u_1 \psi_2 \Big|_{x_2=0}. \label{bc_ferm_1}
\end{equation}

\section{SUSY preserving boundary conditions}

For supersymmetry, it is necessary that the variations of the fields on the boundary are consistent with the boundary conditions. Further, the supersymmetric variation of the action must vanish.

\textit{Dirichlet boundary condition on scalar:} When Dirichlet boundary condition 
\begin{eqnarray}
\phi\Big|_{\partial M}=0\label{bc_scalar_Diri} 
\end{eqnarray}
 is imposed on the scalar, the variation $\delta\phi$  must also vanish on the boundary at $x_2=0$. This requires $\left[\bar{\epsilon} \psi + \bar{\psi} \epsilon \right]_{\partial M}=0$ which yields a relation between the supersymmetry parameters:
\begin{eqnarray}
\epsilon_1 = u_1  \epsilon_2.  \label{cond_Diri} 
\end{eqnarray}
With the SUSY parameter satisfying (\ref{cond_Diri}), it is easy to see that  $\left[\bar{\epsilon} \gamma^2 \psi + \bar{\psi}\gamma^2 \epsilon \right]_{\partial M}=0$ which means $\delta A_2\Big|_{\partial M}=0$. Therefore, supersymmetry with Dirichlet boundary condition on scalar will require imposing  the boundary condition \begin{eqnarray}
A_2\Big|_{\partial M}=0 \label{gauge_cond_A2} 
\end{eqnarray} 
on the gauge field. 

With the gauge field satisfying (\ref{gauge_cond_A2}), the self-adjointness of $\widehat{\mathcal{H}}_G$ demands also imposing the boundary condition (see eqn. (\ref{bc_gauge_2}))
\begin{eqnarray}
\partial_2 A_1\Big|_{\partial M}=0. \label{gauge_cond_normalder_A1} 
\end{eqnarray}
   Therefore, SUSY variation  $\delta(\partial_2 A_1)$ must also vanish on the boundary which gives
\begin{eqnarray}
\left[\bar{\epsilon} \gamma^1 \partial_2\psi + \bar{\psi}\gamma^1 \partial_2\epsilon \right]_{\partial M}=0. 
\end{eqnarray}
The above is satisfied with an additional boundary condition at $x_2=0$
\begin{eqnarray}
(\partial_2 \psi)_+ =  - U_F \gamma^0 (\partial_2 \psi)_-. \label{ferm_cond_2}
\end{eqnarray}
The emergence of this new condition is imperative in a supersymmetric theory:  as the supercharges $Q_\alpha$ obey $\{ Q_\alpha, \bar{Q}_\alpha \}\psi   \propto \widehat{\mathcal{H}}_f \psi$, it is necessary to ensure that $(\widehat{\mathcal{H}}_f \psi)$ is also  in the domain  $\mathcal{D}_{\widehat{ \mathcal{H}}_f}$, else SUSY will change the domain of  $\widehat{ \mathcal{H}}_f$. This is ensured by  (\ref{ferm_cond_2}). 

Further, the SUSY variation of the fermion field on the boundary must satisfy $\delta \psi _+=   U_F \gamma^0  \delta\psi_-$. With boundary conditions (\ref{bc_scalar_Diri}),  (\ref{gauge_cond_A2}) and (\ref{gauge_cond_normalder_A1}) on the scalar and the gauge fields, this is only satisfied if  $u_1 = \pm1$ and the fermionic boundary condition is 
\begin{eqnarray}
\psi_1\Big|_{\partial M} = \pm \psi_2\Big|_{\partial M}. \label{bc_ferm_1} 
\end{eqnarray}

 Under the SUSY transformation, the variation of the action is given by
\begin{eqnarray}
&&\hspace*{-0.7cm} \delta S  = \int_{\partial M} \Big[i\left( \bar{\epsilon} \gamma^\nu  \psi - \bar{\psi} \gamma^\nu \epsilon \right) F_{2 \nu} -\frac{i}{4} \bar{\epsilon} \gamma^{\nu\rho }\gamma^2  \psi F_{ \rho \nu } -   
\nonumber \\ 
&& \quad\quad 
\bar{\psi} \gamma^2 \epsilon  D+  i\bar{\epsilon} \gamma^\nu\gamma^2  \psi \partial_\nu \phi- i(\bar{\epsilon} \psi-\bar{\psi} \epsilon   ) \partial^2 \phi -
 \nonumber \\
&& \quad\quad m  (\bar{\epsilon}\gamma^2 \psi + \bar{\psi} \gamma^2 \epsilon)  \phi\Big]_{x_2=0}.  \label{deltaS}
\end{eqnarray}
and it is straightforward to see that  $\delta S $ vanishes for both $m=0$ and $m \neq 0$ with the boundary conditions (\ref{bc_scalar_Diri} -- \ref{bc_ferm_1}). 

However, unlike the Maxwell-Chern-Simons theory (\ref{Lagrangian_full_space}) on the full (2+1)-dimensional spacetime, here,  the SUSY parameters satisfy (\ref{cond_Diri} ) and as a result there is only $\mathcal{N}=1$ SUSY in the system, generated by two super charges: $Q_D = Q_1 - u_1 Q_2$ and $\bar{Q}_D = \bar{Q}_1 - u_1 \bar{Q}_2$.  

\textit{Neumann boundary condition on scalar:} Imposing Neumann boundary condition on the scalar 
\begin{eqnarray}
\partial_2\phi\Big|_{\partial M}=0\label{bc_scalar_Neumann} 
\end{eqnarray}
 requires the variation $\delta(\partial_2\phi)$  to vanish on the boundary at $x_2=0$, satisfying $\left[\bar{\epsilon} \partial_2 \psi +  \partial_2 \bar{\psi} \epsilon \right]_{\partial M}=0$. With (\ref{ferm_cond_2}), this requires the SUSY parameters to be related as
\begin{eqnarray}
\epsilon_1 =- u_1  \epsilon_2.  \label{cond_Neumann} 
\end{eqnarray}
Using (\ref{cond_Neumann}), it is easy to see that  $\delta A_1\Big|_{\partial M} = \left[\bar{\epsilon} \gamma^1 \psi + \bar{\psi}\gamma^1 \epsilon \right]_{\partial M}=0$. Therefore, supersymmetry with Neumann boundary condition on scalar will require imposing  
\begin{eqnarray}
A_1\Big|_{\partial M}=0 \label{gauge_cond_A1} 
\end{eqnarray} 
on the gauge field.

Again, with the boundary conditions (\ref{bc_scalar_Neumann}) and (\ref{gauge_cond_A1}) on the scalar and gauge fields,  the SUSY variation of the fermion
field satisfy $\delta \psi _+  =   U_F \gamma^0  \delta\psi_-$ only if $u_1 = \pm 1$. Therefore, again,  (\ref{bc_ferm_1}) is the only fermionic boundary condition that  is consistent with SUSY variation of the fields.

In contrary to the previous case, in the variation of action $\delta S$ given in  (\ref{deltaS}), the term $\Big[m  (\bar{\epsilon}\gamma^2 \psi + \bar{\psi} \gamma^2 \epsilon)  \phi\Big]_{x_2=0}$ arising from the variation of the Chern-Simons action, does not vanish with  boundary conditions (\ref{bc_scalar_Neumann} -- \ref{gauge_cond_A1} ) and (\ref{bc_ferm_1}). Therefore, imposing the set of boundary  (\ref{bc_scalar_Neumann} -- \ref{gauge_cond_A1} ) and (\ref{bc_ferm_1}) can only lead to a supersymmetry in a pure Maxwell theory with $m=0$. Further, because the SUSY parameters satisfy the condition (\ref{cond_Neumann}), the $m=0$ case can have  $\mathcal{N}=1$ SUSY  generated by the two supercharges  $Q_N = Q_1 + u_1 Q_2$ and $\bar{Q}_N = \bar{Q}_1 + u_1 \bar{Q}_2$.    In the Maxwell-Chern-Simons theory with $m \neq 0$, SUSY remains completely broken. 


On the other hand,  it is easy to check that the SUSY variations cannot be consistent  with Robin boundary condition $(U_s \neq \pm 1)$ on the scalar, and hence, supersymmetry remains completely broken  if either Dirichlet or Neumann boundary condition is not imposed on the scalar field. This is similar to the findings in \cite{Asorey:2006pr, Asorey:2008xt, Acharyya:2015swa, Asorey:2015lja}.

\begin{table*}
{\small
\begin{tabular}{|c|c|c|c|c|c|} \hline  &&&&& \\
\quad  Boundary  \quad  &  \quad Boundary   \quad & \quad Boundary \quad  &  & &\\
 condition on & condition on  & condition on  & \quad\quad $m=0$ \quad\quad&\quad\quad $m> 0$ \quad\quad &\quad\quad $m< 0$ \quad\quad \\ 
  scalar field  &gauge field &fermion field &&& \\ \hline  &&&&& \\ 
  $ \partial_2\phi \Big|_{x_2 =0} =0$ & $A_1 \Big|_{x_2 =0} =0$  & $ \psi_1\Big|_{x_2=0}  = \pm \psi_2 \Big|_{x_2=0}$ & $\mathcal{N}=1$ SUSY  & No SUSY & No SUSY  \\
 &&&$\epsilon_1 = \mp \epsilon_2$&& \\ &&&&& \\  \hline  &&&&& \\ 
    $\phi \Big|_{x_2 =0} =0$ & $A_2 \Big|_{x_2 =0} =0$  & $ \psi_1\Big|_{x_2=0}  =  \psi_2 \Big|_{x_2=0}$ & $\mathcal{N}=1$ SUSY  & $\mathcal{N}=1$ SUSY & SUSY broken  \\
  &&&&&by edge states \\ & $ \partial_2A_1 \Big|_{x_2 =0} =0$ &&$\epsilon_1 =  \epsilon_2$&$\epsilon_1 =  \epsilon_2$& \\ &&&&& \\   \hline  &&&&& \\ 
      $\phi \Big|_{x_2 =0} =0$ & $A_2 \Big|_{x_2 =0} =0$  & $ \psi_1\Big|_{x_2=0}  =  -\psi_2 \Big|_{x_2=0}$ & $\mathcal{N}=1$ SUSY  & SUSY broken & $\mathcal{N}=1$ SUSY  \\ &&&&by edge states&\\
  & $ \partial_2A_1 \Big|_{x_2 =0} =0$ &&$\epsilon_1 =  -\epsilon_2$&&$\epsilon_1 =  -\epsilon_2$ \\  &&&&& \\  \hline 
\end{tabular}}
\caption{Only combinations of boundary conditions in Maxwell-Chern-Simons theory which can preserve SUSY (partially) } \label{table_}
\end{table*}

\section{ Edge states and  anomalous breaking of SUSY}  
 For non-zero values of $m$, the SUSY variation of the action $\delta  S $ vanishes when the boundary conditions (\ref{bc_scalar_Diri} -- \ref{bc_ferm_1}) are imposed on the fields. One would expect the system to be supersymmetric in such scenarios. However, for supersymmetry in the quantum theory, we need to ensure that the domains of the bosons and the fermions are preserved by the SUSY transformations.  This means, for every fermionic eigenstate of $\widehat{\mathcal{H}}_f$ with eigenvalue $E$, there must be a bosonic energy eigenstate with same eigenvalue (only for states with $E=0$, the pairing is not required \cite{Acharyya:2015swa}). 
 
For $m\neq 0$, there exist counter-propagating modes
 \begin{eqnarray}
&& \psi_{k}^{L} = a_k^{L} \left( \begin{array}{c}
1\\ -\text{sgn}(m) 
\end{array}\right)e^{ i k x_1 - |m| x_2},  \quad\quad
\psi_{k}^{R} = a_k^{R} \left( \begin{array}{c}
1\\ -\text{sgn}(m) 
\end{array}\right)e^{-i k x_1 - |m| x_2} 
\label{edge_states}
\end{eqnarray}
which are eigenstates of  $\widehat{\mathcal{H}}_f$:
\begin{eqnarray}
\widehat{\mathcal{H}}_f \psi_{k}^{L} = E_+ \psi_{k}^{L}, \quad \widehat{\mathcal{H}}_f \psi_{k}^{R} = E_- \psi_{k}^{R}, \quad E_\pm = \pm k
\end{eqnarray}
with fermionic boundary condition
\begin{eqnarray}
\psi_1 \Big|_{x_2=0} = -\text{sgn}(m)\psi_2 \Big|_{x_2=0}. \label{bc_edge_states}
\end{eqnarray}
Here, ${a}_k^{L}$ and ${a}_k^{R}$ are the normalization constants.

For sufficiently large $|m|$, these modes decay exponentially in the bulk $x_2>0$ and are therefore localized near the boundary. 
In contrast, the scalar $\phi$ with Dirichlet boundary condition (\ref{bc_scalar_Diri}) and the gauge fields with boundary conditions (\ref{gauge_cond_A2} -- \ref{gauge_cond_normalder_A1}) have no edge-localized modes. Thus, when the fermionic edge states are present in the domain of $\widehat{\mathcal{H}}_f$ are not paired with any bosonic modes.  This can be elucidated by the fact that the fermionic edge states (\ref{edge_states}) do not satisfy (\ref{ferm_cond_2}) and hence cannot be compatible with the SUSY variation of (\ref{gauge_cond_normalder_A1}). Hence,  in the system with  such fermionic edge states in the domains of $\widehat{\mathcal{H}}_f$, supersymmetry  remains broken despite the invariance of $S$. In other words, the SUSY is anomalous for the fermionic boundary condition (\ref{bc_edge_states}). 

Hence, for non-zero $m$, $\mathcal{N}=1$ SUSY is present only when the fields satisfy the bosonic boundary conditions (\ref{bc_scalar_Diri}-\ref{gauge_cond_normalder_A1}) and fermionic boundary conditions 
\begin{eqnarray}
 \psi_1 \Big|_{x_2=0} =\text{sgn}(m) \psi_2 \Big|_{x_2=0}, \quad\quad  \epsilon_1 =\text{sgn}(m) \epsilon_2. 
\end{eqnarray}
On the other hand, with  bosonic boundary conditions (\ref{bc_scalar_Diri}-\ref{gauge_cond_normalder_A1}) and fermionic boundary conditions (\ref{bc_edge_states})
$\mathcal{N}=1$ supersymmetry  is anomalously broken due to the presence of the fermionic edge states.  

The edge states (\ref{edge_states}) are protected by the mass gap and consequently, there is no such  anomalous SUSY in the massless theory. 

\section{Discussion} 
To summarize, we have demonstrated that in the topologically massive (2+1)-dimensional gauge theory,  supersymmetry cannot emerge with arbitrary choice of boundary conditions.  Only a certain subset of the allowed boundary conditions, which leads to self-adjoint domains of the Hamiltonian, can preserve SUSY (see Table \ref{table_}). Further, we found that even when the action remains invariant under the supersymmetry transformations with certain boundary conditions, there might exist fermionic edge states. Such edge states do not have superpartners and consequently, the  unpaired edge states lead  to anomalous breaking of the supersymmetry in the topologically massive gauge theory.  Consequently, in presence of fermionic edge states, supersymmetry does not exist in the (2+1)-dimensional gauge theory. 

In contrary, a (2+1)-dimensional massive hypermultiplet, which gains relevance in a topological insulators \cite{Grover:2013rc}, can exhibit SUSY with edge states. The hypermultiplet on (2+1)-dimensional manifold $M$ is comprised of complex scalar field $\phi$ and Dirac fermion $\psi$ and SUSY is anomalously broken when  Dirichlet $\phi\Big|_{x_2=0} =0$ or Neumann $\partial_2\phi\Big|_{x_2=0} =0$  is imposed on the scalar and  the fermion satisfies $\psi_1\Big|_{x_2=0} = -\text{sgn}(m) \psi_2\Big|_{x_2=0}$.   On the other hand, imposing Robin boundary condition on the scalar breaks SUSY, in general, for both massless and massive hypermultiplet, except in one  scenario when the scalar satisfies  $\Big[\partial_2\phi + |m| \phi\Big]_{x_2=0}=0$ and the fermionic boundary condition is  $\psi_1\Big|_{x_2=0} = -\text{sgm}(m) \psi_2\Big|_{x_2=0}$. 
In this case, both the scalar modes $\phi_k = c_k^\pm e^{\pm ikx_1-|m|x_2}$  and the fermion modes  (\ref{edge_states}) with energy $E= \pm k$ satisfying the above boundary conditions are  localized near the boundary and decays into the bulk for sufficiently large $|m|$.  Consequently, there is a $\mathcal{N}=1$ SUSY among the edge states with a SUSY parameter sastisfying $\epsilon_1 = \text{sgn}(m) \epsilon_2$. 

In a (1+1)-dimensional Hypermultiplet,  SUSY with the edge states can emerge even with Dirichlet or Neumann boundary conditions on the scalar  as there are fermionic edge states only with $E=0$ \cite{Acharyya:2015swa}.

The analysis presented in this article concerns only $U(1)$ gauge theory and the maximum SUSY in presence of boundaries is $\mathcal{N}=1$.  It is straightforward to generailze to larger multiplets which can lead to extended supersymmetry in the manifolds with boundaries. Further, the analysis can be extended to non-Abelian gauge theories and there too, we expect to find similar results in presence of the topological mass. 


\mbox{}\\
\textbf{Acknowledgements:} We thank Sachindeo Vaidya for valuable discussions and suggestions.



\widetext
\newpage
\appendix

\section{Appendix: Differential forms}
In two dimensions, an one-form $\alpha$ is defined  as
\begin{eqnarray}
\alpha \equiv \alpha_i dx^i = \alpha_1 dx^1 + \alpha_2 dx^2.
\end{eqnarray}
The  exterior derivative gives the two-form:
\begin{eqnarray}
d\alpha = \partial_i \alpha_j dx^i \wedge dx^j = (\partial_1 \alpha_2 -\partial_2 \alpha_1) dx^1 \wedge dx^2
\end{eqnarray}
which is the volume form. 

The Hodge-$\ast$ operation 
 is defined using
\begin{eqnarray}
\ast dx^1 =- dx^2, \quad\quad  \ast dx^2 =  dx^1, \quad\quad   \ast(dx^1 \wedge dx^2) =1, \quad\quad \ast 1= dx^1 \wedge dx^2. 
\end{eqnarray}

Therefore,  $\ast \alpha$ is an one-form and  $\ast d\alpha$ is a 0-form:
\begin{eqnarray}
&& \ast \alpha =  \alpha_2 dx^1 - \alpha_1 dx^2, \quad\quad \ast d\alpha =  (\partial_1 \alpha_2 -\partial_2 \alpha_1).
\end{eqnarray}
From above, it is easy to see that $ (d \ast d\alpha)$ and ($\ast  d \ast d\alpha$) are both one-forms: 
\begin{eqnarray}
&& d \ast d\alpha =  \partial_1(\partial_1 \alpha_2 -\partial_2 \alpha_1)dx^1 +  \partial_2(\partial_1 \alpha_2 -\partial_2 \alpha_1)dx^2, \\
&& \ast d \ast d\alpha =  \partial_2(\partial_1 \alpha_2 -\partial_2 \alpha_1)dx^1 - \partial_1(\partial_1 \alpha_2 -\partial_2 \alpha_1)dx^2. 
\end{eqnarray}

In Hodge theory, 
the Hilbert space is defined using the inner product 
\begin{eqnarray}
\langle \beta, \alpha \rangle \equiv \int_M dx_1 dx_2 \,\,  \beta_i^\dagger \alpha_i.
\end{eqnarray}

Therefore, 
\begin{eqnarray}
\langle\beta,  \ast d \ast d\alpha \rangle &=&  \int_M dx_1 dx_2 \,\,\left(\beta_1^\dagger  \partial_2(\partial_1 \alpha_2 -\partial_2 \alpha_1)  -   \beta_2^\dagger \partial_1(\partial_1 \alpha_2 -\partial_2 \alpha_1)\right) \\
&=&   \int_M dx_1 dx_2 (\partial_1 \beta_2^\dagger - \partial_2  \beta_1^\dagger)(\partial_1 \alpha_2 -\partial_2 \alpha_1) + \int_{\partial M} dx_1 \beta_1 (\partial_1 \alpha_2 -\partial_2 \alpha_1)\Big|_{x_2=0}  \\ 
&=&\int_M dx_1 dx_2 (\ast d\beta)^\dagger (\ast d\alpha) + \int_{\partial M} dx_1 \beta_1 (\partial_1 \alpha_2 -\partial_2 \alpha_1)\Big|_{x_2=0}.
\end{eqnarray}


\thebibliography{01}


\bibitem{Witten:1988hf}
E.~Witten,
Commun. Math. Phys. \textbf{121}, 351-399 (1989)


\bibitem{Elitzur:1989nr}
S.~Elitzur, G.~W.~Moore, A.~Schwimmer and N.~Seiberg,
Nucl. Phys. B \textbf{326}, 108-134 (1989)


\bibitem{Wen:1995}
X.-G. Wen, 
Adv. Phys. {\bf 44}, 405 (1995).


\bibitem{Grivin:1987}
S. M. Girvin and A.H. MacDonald, 
Phys. Rev. Lett. 58, 1252 (1987)

\bibitem{Zhang:1988}
S.C. Zhang, T.H. Hansson and S. Kivelson, 
Phys. Rev. Lett. 62,82 (1988).


\bibitem{Laughlin:1988} 
R. B. Laughlin, 
 Science 242, 525 (1988)
 
 



\bibitem{Kalmeyer:1987}
V. Kalmeyer and R. B. Laughlin, 
Phys. Rev. Lett. 59, 2095 (1987).

\bibitem{Kou:2008} 
S.-P. Kou, M. Levin, and X.-G. Wen, 
Phys. Rev. B 78, 155134 (2008).

\bibitem{Hansson:2004}
 T. H. Hansson, V. Oganesyan, and S. L. Sondhi, 
Ann. Phys. (NY) 313, 497 (2004).

\bibitem{Palumbo:2013rb}
G.~Palumbo and J.~K.~Pachos,
Phys. Rev. Lett. \textbf{110}, no.21, 211603 (2013)


\bibitem{Mechelen:2020}
T.V. Mechelen and Z. Jacob, 
Phys. Rev. B {\bf 102}, 155425 (2020)


\bibitem{Grover:2013rc}
T.~Grover, D.~N.~Sheng and A.~Vishwanath,
Science \textbf{344}, no.6181, 280-283 (2014)

\bibitem{Rahmani:2015}
A. Rahmani, X. Zhu, M. Franz, and I. Affleck, 
Phys. Rev. Lett., {\bf 115}, no. 16, 66401 (2015)


\bibitem{Hsieh:2016emq}
T.~H.~Hsieh, G.~B.~Hal\'asz and T.~Grover,
Phys. Rev. Lett. \textbf{117}, no.16, 166802 (2016)


\bibitem{Jian:2016zll}
S.~K.~Jian, C.~H.~Lin, J.~Maciejko and H.~Yao,
Phys. Rev. Lett. \textbf{118}, no.16, 166802 (2017)


\bibitem{Prakash:2020krs}
A.~Prakash and J.~Wang,
Phys. Rev. B \textbf{103}, no.8, 085130 (2021)


\bibitem{Ma:2021dua}
K.~K.~W.~Ma, R.~Wang and K.~Yang,
Phys. Rev. Lett. \textbf{126}, no.20, 206801 (2021)


\bibitem{Bae:2021lvk}
J.~B.~Bae and S.~Lee,
SciPost Phys. \textbf{11}, no.5, 091 (2021)


\bibitem{Belyaev:2008xk}
D.~V.~Belyaev and P.~van Nieuwenhuizen,
JHEP \textbf{04}, 008 (2008)

\bibitem{Okazaki:2013kaa}
T.~Okazaki and S.~Yamaguchi,
Phys. Rev. D \textbf{87}, no.12, 125005 (2013)


\bibitem{Acharyya:2015swa}
N.~Acharyya, M.~Asorey, A.~P.~Balachandran and S.~Vaidya,
Phys. Rev. D \textbf{92}, no.10, 105016 (2015)






\bibitem{Esteve:1986db}
J.~G.~Esteve,
Phys. Rev. D \textbf{34}, 674-677 (1986)






\bibitem{Hook:2013yda}
A.~Hook, S.~Kachru and G.~Torroba,
JHEP \textbf{11}, 004 (2013)

\bibitem{Asorey:2006pr}
M.~Asorey, D.~Garcia-Alvarez and J.~M.~Munoz-Castaneda,
J. Phys. A \textbf{39}, 6127-6136 (2006)

\bibitem{Asorey:2008xt}
M.~Asorey and J.~M.~Munoz-Castaneda,
J. Phys. A \textbf{41}, 304004 (2008)

\bibitem{Balachandran:1991dw}
A.~P.~Balachandran, G.~Bimonte, K.~S.~Gupta and A.~Stern,
Int. J. Mod. Phys. A \textbf{7}, 4655-4670 (1992)

\bibitem{Balachandran:1993tm}
A.~P.~Balachandran, L.~Chandar, E.~Ercolessi, T.~R.~Govindarajan and R.~Shankar,
Int. J. Mod. Phys. A \textbf{9}, 3417-3442 (1994)






\bibitem{Asorey:2004kk}
M.~Asorey, A.~Ibort and G.~Marmo,
Int. J. Mod. Phys. A \textbf{20}, 1001-1026 (2005)


\bibitem{Balachandran:1994vi}
A.~P.~Balachandran, L.~Chandar and E.~Ercolessi,
Int. J. Mod. Phys. A \textbf{10}, 1969-1993 (1995)


\bibitem{Acharyya:2016xaq}
N.~Acharyya, A.~P.~Balachandran, V.~Errasti D\'\i{}ez, P.~N.~Bala Subramanian and S.~Vaidya,
Phys. Rev. D \textbf{94}, no.8, 085026 (2016)








\bibitem{Asorey:2013wvh}
M.~Asorey, A.~P.~Balachandran and J.~M.~P\'erez-Pardo,
JHEP \textbf{12}, 073 (2013)

\bibitem{Asorey:2015sra}
M.~Asorey, A.~P.~Balachandran and J.~M.~Perez-Pardo,
Rev. Math. Phys. \textbf{28}, no.09, 1650020 (2016)


\bibitem{Asorey:2015lja}
M.~Asorey, D.~Garc\'\i{}a-Alvarez and J.~M.~Mu\~noz-Casta\~neda,
Int. J. Geom. Meth. Mod. Phys. \textbf{12}, no.06, 1560004 (2015)
\end{document}